\documentclass[12pt,a4paper]{article}
\usepackage[pdfstartview=FitH,colorlinks=true,linkcolor=black,anchorcolor=black,citecolor=black,urlcolor=black]{hyperref}
\usepackage[english]{babel}
\usepackage[top=2.7cm,right=2.3cm,left=2.3cm,bottom=2.7cm]{geometry}
\usepackage{amsmath,amssymb,titling,authblk}
\usepackage{amsmath,slashed}
\usepackage{cite}
\usepackage{xcolor}

\allowdisplaybreaks[4]
\numberwithin{equation}{section}

\newcommand{\be}{\begin{equation}}
\newcommand{\ee}{\end{equation}}
\newcommand{\bea}{\begin{eqnarray}}
\newcommand{\eea}{\end{eqnarray}}
\newcommand{\ii}{\mathrm{i}}
\newcommand{\e}{\mathrm{e}}
\newcommand{\dd}{\mathrm{d}}

\usepackage{color}

\newenvironment{remark}[1][Remark]{\begin{trivlist}
\item[\hskip \labelsep {\bfseries #1}]\small\sl }{\end{trivlist}}

\begin{document}

\setlength{\droptitle}{-6pc}
\pretitle{\begin{flushright}\small
ICCUB-16-019
\end{flushright}\vspace*{2pc}%
\begin{center}\LARGE}
\posttitle{\par\end{center}}

\title{Wick Rotation and Fermion Doubling in\\[3pt] Noncommutative Geometry\vspace{5pt}}

\renewcommand\Affilfont{\itshape}
\setlength{\affilsep}{1.5em}
\renewcommand\Authands{ and }

\author[1,2]{Francesco D'Andrea} 
\author[3]{Maxim A.~Kurkov}
\author[2,4,5]{Fedele Lizzi}
\affil[1]{Dipartimento di Matematica e Applicazioni ``Renato Caccioppoli", \mbox{Universit\`{a} di Napoli {\sl Federico II}}\vspace{5pt}}
\affil[2]{INFN, Sezione di Napoli\vspace{5pt}}
\affil[3]{
CMCC-Universidade Federal do ABC, Santo Andr\'e, S.P., Brazil
\vspace{5pt}}
\affil[4]{Dipartimento di Fisica ``Ettore Pancini'', Universit\`{a} di Napoli {\sl Federico II}\vspace{5pt}}
\affil[5]{Departament de Estructura i Constituents de la Mat\`eria, Institut \mbox{de Ci\'encies del Cosmos, Universitat de Barcelona}}

\date{}

\maketitle 

\begin{abstract}\noindent
In this paper, we discuss two features of the noncommutative geometry and spectral action approach to the Standard Model: the fact that the model is inherently Euclidean, and that it requires a quadrupling of the fermionic degrees of freedom. We show how the two issues are intimately related. We give a precise prescription for the Wick rotation from the Euclidean theory to the Lorentzian one, eliminating the extra degrees of freedom. This requires not only projecting out mirror fermions, as has been done so far, and which leads to the correct Pfaffian, but also the elimination of the remaining extra degrees of freedom. The remaining doubling has to be removed in order to recover the correct Fock space of the physical (Lorentzian) theory. In order to get a Spin(1,3) invariant Lorentzian theory from a Spin(4) invariant Euclidean theory such an elimination must be performed \emph{after} the Wick rotation.
Differences between the Euclidean and Lorentzian case are described in detail, in a pedagogical way.
\end{abstract}

\newpage


\section{Introduction}
Noncommutative geometry~\cite{Connesbook, Landi, ticos}
generalizes some notions and tools from differential geometry to the study of quantum spaces, ``geometric'' objects that are described by (noncommutative) operator algebras.
It is based on results valid for (compact) Riemannian manifolds, and by its nature not immediately suited to accommodate a Lorentzian signature of the space. Although there are attempts in this direction --- using either Krein spaces~\cite{Strohmaier:2001zx,PaschkeSitarz, vandendungen, BrouderBiziBesnard}, covariant approaches \cite{PaschkeVerch}, {Wick rotations on pseudo-Riemanninan structures~\cite{KoenMarioAdam}},or algebraic characterizations of causal structures~\cite{Moretti, FrancoEckstein, BesnardBizi} ---
it is fair to say that we are still far away from a full understanding of the theory.
This becomes a problem when the tools of noncommutative geometry are applied to physics, and in particular to the Standard Model via the spectral action~\cite{ChamseddineConnesspectralaction, AC2M2, Walterbook}.
The theory is now reaching a sufficient level of maturity to be compared with phenomenology, but in order to do this, as explained e.g.~in \cite[p.~218]{CM08}, one has to start with an Euclidean theory ``leaving as an important problem the Wick rotation back to the Minkowski signature'' (cit.).
The starting point is an action functional defined in a purely spectral fashion from a suitable almost commutative space (a product of a manifold and a matrix geometry). On one side this procedure allows to reproduce several features of the Standard Model, not only qualitatively, but quantitatively as well.
On the other hand, the Wick rotation in this context requires some clarifications.

There is another feature in the noncommutative geometry approach to the Standard Model, which makes a comparison with phenomenology not completely straightforward.
It is the so-called ``fermion doubling''~\cite{LMMS,G-BIS}, although it consists in fact in a quadruplication of the degrees of freedom. 

In the spectral action approach the Hilbert space of the theory is a product of two factors. One is given by Dirac spinors on an ordinary $4$-dimensional manifold, locally given by $4$ complex valued functions. The finite space is basically $\mathbb C^N$, where $N$ equals to number of particles and antiparticles in the Standard Model: lepton left doublet, two leptonic singlets, the same for quarks times three colors, this makes 16, times two for antiparticles, times three generations. In the end we get $N = 96$.
The full Hilbert space is given, locally, by vector-valued functions with $4N=384$ components, four times what expected from physics.
Perhaps the most dangerous part of such a quadruplication, is the presence of mirror fermions, i.e.\  fermions with the same (gauge) quantum numbers as the original ones (hypercharge, isospin, color), but opposite chirality. The remaining doubling is related with the fact, that in this approach the spinor multiplets with quantum numbers of particles and antiparticles enter in the Lagrangian as independent fields, not conjugated to each other. 

A Lorentzian version of the Standard Model's spectral action was presented by Barrett in~\cite{Barrett}. There he recasts the spectral data of the noncommutative geometry approach to the Standard Model in Lorentzian form, and discusses how to deal with the fermion doubling problem.
The present paper has considerable connections with Barrett's work, but our point of view is in the construction of full fledged Euclidean theory. This is necessary since, as we said, it is not clear yet how to generalize noncommutative geometry to Lorentzian signature.

A slightly different solution of the mirror fermion doubling was offered in~\cite{AC2M2}: the fermionic action
proposed there depends on $2N=192$ independent complex valued functions. This solves the problem at the level of the fermionic functional integral, but not of the full quantum field theory, which requires the construction of the physical Fock space via canonical quantization. A peculiarity of Grassmann integral is that the Pfaffian is insensitive to the presence of the
remaining doubling (see Appendix~\ref{apppathintegral}). 
On the other hand, the Fock space construction via canonical quantization (which has to be carried out after Wick rotation), 
is sensitive to such a doubling (see Sect.~\ref{se:genralpres} for discussions). We need then a prescription to eliminate the remaining extra degrees of freedom in order to obtain, strictly speaking, the Standard Model.
 
The passage from the Euclidean
action functional from noncommutative geometry to its Lorentzian version has never been done and understood in detail. The aim of this paper is to give a coherent and detailed prescription\footnote{{For the bosonic spectral action we consider a local structure given by a finite number of terms of the proper asymptotic expansion.}} for this procedure, accompanied by the elimination of the extra degrees of freedom.
We will argue that
the passage to a Lorentzian signature  must be done first, in order to start with a Spin(4) invariant Euclidean action and get a Spin(1,3) invariant Lorentzian action. From another side the presence of extra degrees of freedom simplifies the Wick rotation procedure,
in particular no modification of the inner product is needed in order to get Spin(1,3) invariant expression from the Spin(4) invariant one.
As a minor remark, we also notice that the procedure of Wick rotation based on imaginary time, which is commonly used in this context (see e.g.~\cite{Schucker}) does not work on a curved spacetime: instead, one has to Wick rotate the vierbeins.

We will present a procedure to pass from the Euclidean theory (motivated by noncommutative geometry) to a Lorentzian one which satisfies the following requirements:
\begin{list}{$\bullet$}{\leftmargin=1.8em}
\item The Euclidean action for bosons, i.e.\ the one using the Euclidean metric tensor $g_{\mu\nu}^{{\rm E}}$ with signature $\left\{+,+,+,+\right\}$, transforms into the correct Lorentzian actions with metric tensor $g_{\mu\nu}^{{\rm M}}$ with signature $\left\{+,-,-,-\right\}$. By ``correct'' we mean the one used in physics, in particular with correct signs in all terms (specifically, the kinetic energy).

\item The Euclidean fermionic action must transform into the correct (acceptable for canonical quantization) Lorentzian fermionic action which appears in the Standard Model.
 
\item 
The quadrupling of degrees of freedom must be eliminated.
\end{list}

The paper is organized  as follows. In Sect.~\ref{se:bosonic} we argue that the proper procedure for the Wick rotation is to rotate the vierbeins rather than the coordinates, providing the reader with all needed technical details, and we discuss the bosonic part of the action functional.
In Sect.~\ref{se:fermionic}, we summarize all delicate points concerning fermions in the contexts of the quadrupling,  and discuss relevant aspects of the interplay between Euclidean and Lorentzian invariance. In Sect.~\ref{se:rotatefermions} we propose and discuss step by step a prescription for the Wick rotation of the fermionic part, with subsequent elimination of extra degrees of freedom.
Sect.~\ref{sec:concl} contains the conclusions. Relevant aspects of path integrals, notations and computational details are collected in the appendices.

\section{Wick rotation: bosonic case \label{se:bosonic}}
Wick rotation is usually performed by rotating the zeroth (time) time coordinate to imaginary values:
\begin{equation}
t \rightarrow \ii  \, t \label{it}. 
\end{equation}
This is well described in the context of noncommutative geometry in~\cite{Schucker}.
The Euclidean and Lorentzian actions are transformed into each other by a Wick rotation\footnote{Greek indexes
$\mu$, $\nu$ run from 0 to 3 in both Euclidean and Lorentzian (curved) cases. The flat case indices $A$, $B$ are raised and lowered using the flat metric, either $\delta = {\rm diag\,}(+1,+1,+,1+,1)$ or
$\eta = {\rm diag\,}(+1,-1,-1,-1)$ depending on the signature. Vierbeins are denoted $e_\mu^A$. When necessary the superscripts ``E'' (Euclidean) and ``M'' (Minkowkian) will be used to distinguish between the Euclidean and Lorentzian cases. Latin indices $i,j$ run from 1 to 3.}
\begin{equation}
\exp{\left(-S^{\rm E}[{\rm fields}, g_{\mu\nu}^{\rm E}] \right)} \longleftrightarrow \exp{\left(\ii S^{\rm M}[{\rm fields}, g_{\mu\nu}^{\rm M}] \right)} \;, \label{cond}
\end{equation}
where ``fields'' generically represents all (fermionic and bosonic) fields present in the theory. The expression \eqref{cond} should then be integrated 
over all fields.

This procedure is  not suitable in general for curved space-time. In~\cite{Visser}, for example, it is explicitly shown that, for different choices of coordinates, the De Sitter metric (which has Lorentzian
signature) transforms in radically different ways. In particular closed, open and flat slicing of the manifold give Euclidean, Lorentzian or even imaginary metric tensors.
This illustrates that, generally speaking, for a coordinate-dependent metric tensor the naive prescription~\eqref{it} does not satisfy the condition~\eqref{cond}.
In particular, unacceptable imaginary kinetic terms can appear. A more robust  prescription, which respects the condition~\eqref{cond}, is to \emph{Wick rotate the vierbeins}. Namely, to pass from the Euclidean to a Lorentzian theory, each expression $F$ which depends on vierbeins has to be transformed according to the rule\footnote{In the case we are interested in, the Euclidean Lagrangian and the volume form are polynomial or at most rational functions of the vierbeins and their derivatives, see in particular~\eqref{eq:volume}, so the prescription is well-defined.}:
\be
\mbox{Wick:} \quad F\left[e_{\mu}^0, e_{\mu}^j \right]
\longrightarrow \quad F\left[\ii e_{\mu}^0, e_{\mu}^j\right],
\,\, \, j = 1,2,3. 
\label{bosWickvier}
\ee
Note that the vierbeins $e_{\mu}^A$, which appear in both sides of the correspondence, are the same real functions.
As customary, we assume that $e^0=e^0_\mu dx^\mu$ is globally defined (time-like after rotation), so that the Wick rotation is well-defined.
The correspondence \eqref{bosWickvier} is obviously invertible,
and the inverse correspondence will be denoted by $\mbox{Wick}^*$:\,\footnote{Usually by ``Wick rotation'' is meant the map from the Lorentzian to the Euclidean theory, here denoted by $\mbox{Wick}^*$. For the scope of the present paper, our terminology is preferable.}
\be
\mbox{Wick}^*:\quad F\left[e_{\mu}^0, e_{\mu}^j \right]
\longrightarrow \quad F\left[-\ii e_{\mu}^0, e_{\mu}^j\right]
\,\, \, j = 1,2,3. 
\ee
In what follows we apply the transformation \eqref{bosWickvier} to the bosonic action $S^{\rm E}_{\rm bos}$, derived from noncommutative geometry. In the Euclidean bosonic action the vierbeins enter only via the metric tensor $g_{\mu\nu}^{\rm E}$, given by:
\begin{equation}
g_{\mu\nu}^{\rm E} = e^{A}_{\mu}e^{B}_{\nu}\delta_{AB}. \label{geucl}
\end{equation}
One can easily see that, applying the Wick rotation~\eqref{bosWickvier} to the metric tensor one gets:
\be
\mbox{Wick:} \quad g_{\mu\nu}^{\rm E} \longrightarrow -g_{\mu\nu}^{\rm M}, \label{GW}
\ee
where
\begin{equation}
g_{\mu\nu}^{\rm M} = e^{A}_{\mu}e^{B}_{\nu}\eta_{AB}.  \label{gmink}
\end{equation}
The volume measure deserves a special comment, since it is the only part in the action which is not a rational function of the metric or its derivatives. We assume, of course, to start with a oriented Riemannian manifold, so that the volume form is defined. If the manifold is oriented, we can choose the vierbeins so that $\det(e^A_\mu)>0$ at every point and in every chart. For an arbitrary pseudo-Riemannian metric $g$,
\begin{equation}\label{eq:volume}
\left|\det g\right|=(\det (e^A_\mu))^2 \;.
\end{equation}
Thus:
\be
\mbox{Wick:} \quad  \sqrt{g^{\rm E}} = \det{\left(e^A_{\mu}\right)}
\longrightarrow  \ii \det{\left(e^A_{\mu}\right)} =  \ii  \sqrt{-g^{\rm M}}  \label{WSR}
\ee
where with a slight abuse of notation we denote by $g^{\rm E}$ and $g^{\rm M}$ the determinant of the respective matrices.

Summarizing, we arrive at the following transformation law for the action, as a functional of the metric:
\begin{align}
\mbox{Wick:} & \quad S^{\mathrm{E}}_{\mathrm{bos}}\left[\mathrm{fields},g_{\mu\nu}^{\mathrm{E}}\right]=\int \mathrm{d}^4x\,
\sqrt{g^{\mathrm{E}}}\,\mathcal{L}^{\mathrm{E}}_{\mathrm{bos}}(\mathrm{fields},g_{\mu\nu}^{\mathrm{E}})
\longrightarrow \notag \\
& \longrightarrow
\mathrm{i}\int \mathrm{d}^4x\,\sqrt{-g^{\mathrm{M}}}\,\mathcal{L}^{\mathrm{E}}_{\mathrm{bos}}(\mathrm{fields},-g_{\mu\nu}^{\mathrm{M}})
\equiv
-\mathrm{i}S^{\mathrm{M}}_{\mathrm{bos}}\left[\mathrm{fields},g_{\mu\nu}^{\mathrm{M}}\right]
\notag \\ &\equiv
-\mathrm{i}\int \mathrm{d}^4x\,
\sqrt{-g^{\mathrm{M}}}\,\mathcal{L}^{\mathrm{M}}_{\mathrm{bos}}(\mathrm{fields},g_{\mu\nu}^{\mathrm{M}})
\end{align}
where we put a $-\mathrm{i}$ factor in front of $S^{\mathrm{M}}_{\mathrm{bos}}$ in order to get the correspondence \eqref{cond}.

Since we are interested in the spectral action, we will consider the dependence on the metric, as well as vector fields, generically indicated by $A_\mu$, and scalar fields (which include the Higgs), generically indicated as $\phi$. 

The (Euclidean) spectral action is a regularized trace of the Dirac operator. The regularization was originally made by considering a cutoff~\cite{ChamseddineConnesspectralaction}, but a $\zeta$-function regularization is also possible~\cite{WKLS}. In either case the contribution involves three terms:
\begin{equation} 
S_{\rm bos}^{\rm E}\left[g_{\mu\nu}^{ \rm E}, A_{\mu}, \phi\right]
 =  S^{\rm E}_{\rm grav}\left[g_{\mu\nu}^{ \rm E}\right]
 +  S^{\rm E}_{\rm gauge}\left[g_{\mu\nu}^{ \rm E}, A_{\mu}\right]
 +  S^{\rm E}_{\rm scal}\left[g_{\mu\nu}^{ \rm E}, A_\mu, \phi\right]. \label{SB}
\end{equation}
where $S^{\rm E}_{\rm grav}$ is purely gravitational, $S^{\rm E}_{\rm gauge}$ is the gauge bosons' action, $S^{\rm E}_{\rm scal}$ is the scalar action. 

We will now be more specific and discuss the three contributions. We illustrate our prescription
for the first three nontrivial heat kernel coefficients, sufficient to recover the Standard Model. Higher coefficients, leading to higher derivative theories, can easily be elaborated in a similar fashion. 
\subsubsection*{Gravitational sector}
The gravitational part of the action is
\begin{equation}
S^{\rm E}_{\rm grav}\left[g_{\mu\nu}^{ \rm E}\right] = \int d^4 x \sqrt{g^{\rm E}} \left(\lambda  +  \frac{M_{\rm Pl}^2}{16\pi}R\left[g_{\mu\nu}^{\rm E}\right] 
+ a  C_{\mu\nu\alpha\beta}\left[g_{\mu\nu}^{\rm E}\right] C^{\mu\nu\alpha\beta}\left[g_{\mu\nu}^{\rm E}\right]\right), \label{gra}
\end{equation}
where $\lambda$ is the cosmological term, $M_{\rm Pl}$ the Planck mass,  $a$ a dimensionless constant,  and $C$  the Weyl tensor. We denote through $R_{\mu\nu\alpha\beta}\left[g_{\mu\nu}\right]$, 
$R_{\mu\nu}\left[g_{\mu\nu}\right]$ and $R\left[g_{\mu\nu}\right]$ correspondingly the Riemann and Ricci tensors and the scalar curvature built from the metric tensor $g_{\mu\nu}$, see the explicit expressions in the Appendix~\ref{appnotation}, where notations and useful formulas are collected.
 Using~\eqref{RMN}, \eqref{RICCI}, \eqref{RS} and  \eqref{W2}, one finds, that
the various terms which enter in the gravitational action \eqref{gra}
transform as:
\begin{align*}
\mbox{Wick}: &&
R\left[  g_{\mu\nu}^{\rm E} \right]
&\longrightarrow
R\left[ - g_{\mu\nu}^{\rm M} \right] = -R\left[  g_{\mu\nu}^{\rm M} \right] , \\[5pt]
\mbox{Wick}: && \hspace*{-7pt}
C_{\mu\nu\alpha\beta} \left[  g_{\mu\nu}^{\rm E} \right]C^{\mu\nu\alpha\beta}\left[  g_{\mu\nu}^{\rm E} \right]
&\longrightarrow
C_{\mu\nu\alpha\beta} \left[  -g_{\mu\nu}^{\rm M} \right] C^{\mu\nu\alpha\beta}\left[ - g_{\mu\nu}^{\rm M} \right] = C_{\mu\nu\alpha\beta} \left[  g_{\mu\nu}^{\rm M} \right]C^{\mu\nu\alpha\beta}\left[  g_{\mu\nu}^{\rm M} \right] .
\end{align*}
Thus
\begin{equation}
\mbox{Wick}:\quad\exp{\left( - S_{\rm grav}^{\rm E} \left[g_{\mu\nu}^{ \rm E}\right]\right)}
 \longrightarrow \exp{\left( \ii S_{\rm grav}^{\rm M} \left[g_{\mu\nu}^{ \rm M}\right]\right)}, 
\end{equation}
where
\begin{equation}
 S_{\rm grav}^{\rm M} \left[g_{\mu\nu}^{ \rm M}\right]=
\int d^4 x \sqrt{-g^{\rm M}} \left(-\lambda \, + \, \frac{M_{\rm Pl}^2}{16\pi}R\left[g_{\mu\nu}^{\rm M}\right] 
- a  C_{\mu\nu\alpha\beta}\left[g_{\mu\nu}^{\rm M}\right] C^{\mu\nu\alpha\beta}\left[g_{\mu\nu}^{\rm M}\right]\right) .
\end{equation}

\subsubsection*{Gauge sector}
The gauge action is
$$
S^{\rm E}_{\rm gauge} =  \int d^4 x \sqrt{g^{\rm E}} \, g^{\mu\alpha}_{\rm E} g^{\nu\beta}_{\rm E} \,{\rm tr}\, F_{\mu\nu} F_{\alpha\beta}.
$$
According to the prescription \eqref{bosWickvier} we obtain
\begin{equation}
\mbox{Wick}:\quad \exp{\left( - S_{\rm gauge}^{\rm E} \left[g_{\mu\nu}^{ \rm E}\right]\right)}
 \longrightarrow \exp{\left( \ii S_{\rm gauge}^{\rm M} \left[g_{\mu\nu}^{ \rm M}\right]\right)}, 
\end{equation}
where
\begin{eqnarray}
 S_{\rm gauge}^{\rm M} \left[g_{\mu\nu}^{ \rm M}, A_{\mu}\right]  = \int d^4 x  \sqrt{-g^{\rm M}} \,\left(- g^{\mu\alpha}_{\rm M} g^{\nu\beta}_{\rm M} \,{\rm tr}\, F_{\mu\nu} F_{\alpha\beta}\right),
 \end{eqnarray}
and again we reproduce the correct action, see e.g.~\cite{PeskinSchroeder}.

\subsubsection*{Scalar sector}
The typical action for a generic scalar multiplet $\phi_j$, $j = 1\ldots N$  
like the Higgs field $H$, is
\begin{equation}
S_{\rm scal}^{\rm E}\left[g_{\mu\nu},A_\mu,  \phi\right] = \int d^4 x \sqrt{g^{\rm E}}
\left\{ \sum_{j=1}^N \left(g^{\mu\nu}_{\rm E} \nabla_{\mu} \phi_j^{\dagger} \nabla_{\nu}\phi_j  -  \frac{1}{6}R\left[g_{\mu\nu}^{\rm E}\right] 
\phi_j^{\dagger} \phi_j\right) + V(\phi)
\right\}, \label{scaleucl}
\end{equation}
The covariant derivatives $\nabla_\mu=\partial_\mu+\ii A_\mu$
contains just  gauge fields, and the potential $V$ does not depend on the metric tensor. 

Applying the transformation \eqref{bosWickvier} to the scalar action \eqref{scaleucl} we immediately obtain:
\begin{equation}
\mbox{Wick}:\quad\exp{\left( - S_{\rm scal}^{\rm E} \left[g_{\mu\nu}^{ \rm E}\right]\right)}
 \longrightarrow \exp{\left( \ii S_{\rm scal}^{\rm M} \left[g_{\mu\nu}^{ \rm M}\right]\right)}, 
\end{equation}
where
\begin{equation}
 S_{\rm scal}^{\rm M} \left[g_{\mu\nu}^{ \rm M}, \phi_j\right]  = \int d^4 x \sqrt{-g^{\rm M}}
\left\{ \sum_{j=1}^N \left(g^{\mu\nu}_{\rm M} \nabla_{\mu} \phi_j^{\dagger} \nabla_{\nu}\phi_j  -  \frac{1}{6}R\left[g_{\mu\nu}^{\rm M}\right] 
\phi_j^{\dagger} \phi_j\right) - V(\phi)
\right\},
\end{equation}
again in agreement with the literature.

We stress that this procedure is valid both for the heat kernel expansion of the spectral action, and for the  resummation introduced in~\cite{BarvisnkiVilkoviski} by Barvinsky and Vilkovisky, and applied to noncommutative geometry in \cite{Vassilievich,KuLiVa}, at least when only a finite number of terms in the expansion are considered.

\section{Fermions \label{se:fermionic}}
The fermionic case is subtle in field theory, the straightforward Wick rotation must be supplemented by other considerations. Moreover, in our case it is necessary to threat properly the extra degrees of freedom due to the fermionic quadrupling. In this section we will describe in detail the fermionic quadrupling and the elimination of mirror degrees of freedom done so far \cite{AC2M2}, as a preparation to the Wick rotation from Euclidean to Lorentzian signature, accompanied by the elimination of the remaining extra degrees of freedom, performed in the next section.  
First we discuss briefly the difference between Euclidean and Lorentzian fermionic theories, focusing on transformation properties and charge conjugation.

\subsection{Spin(4) vs Spin(1,3)}
In the rotation from an Euclidean theory to a Lorentzian one, the symmetries of the theory go from Spin(4), the universal covering of SO(4), to Spin(1,3), which covers the Lorentz group.
Let us first fix notations. We work in chiral basis in which
\begin{equation}
\gamma^5 = \left(
\begin{array}{cc}
-\sigma_0 & 0_{2\times2} \\
0_{2\times2} & \sigma_0 
\end{array}
\right), \quad \gamma^0_{\rm E} =  \left(
\begin{array}{cc}
0_{2\times2} & \sigma^0   \\
\sigma^0 & 0_{2\times2} 
\end{array}
\right), \quad \gamma^j_{\rm E} =  \left(
\begin{array}{cc}
0_{2\times2} & -\ii\sigma^j   \\
\ii\sigma^j & 0_{2\times2} 
\end{array}
\right),
\end{equation}
where $\sigma^j$ are the Pauli matrices and $\sigma^0$ is the two by two unity matrix.
In particular the anticommutator of the Euclidean gamma matrices reads:
\begin{equation}
\left\{ \gamma^A_{\rm E},\gamma^B_{\rm E} \right\} = 2 \delta^{AB}. \label{gammaE}
\end{equation}
The matrix $\gamma^5$ is the product of four Euclidean Dirac matrices
\begin{equation}
\gamma^5 = \gamma^0_{\rm E}\gamma^1_{\rm E}\gamma^2_{\rm E}\gamma^3_{\rm E}
= \ii \gamma^0_{\rm M}\gamma^1_{\rm M}\gamma^2_{\rm M}\gamma^3_{\rm M},
\end{equation}
where the Lorentzian Dirac matrices in the same basis are defined  by
\begin{equation}
\gamma^{0}_{\rm M} \equiv  \gamma^{0}_{\rm E}, \quad \gamma^{j}_{\rm M} \equiv  \ii\gamma^{j}_{\rm E},\,\, j=1,2,3  \label{gammaEM}
\end{equation}
in agreement with \eqref{gammaE} and
\begin{equation}
\left\{ \gamma^A_{\rm M},\gamma^B_{\rm M} \right\} = 2 \eta^{AB}. \label{gammaM}
\end{equation}
It is also convenient to rewrite the Lorentzian Dirac matrices
defined by \eqref{gammaEM} in the following form, which we will use in Sect.~\ref{se:rotatefermions}
\begin{equation}
\gamma^{A}_{\rm M} = \left(\begin{array}{cc}
0_{2\times2} & \sigma^{A} \\
\bar{\sigma}^A & 0_{2\times 2}
\end{array}\right), \label{gammaMM}
\end{equation}
where
\begin{equation}
\sigma \equiv \left\{\sigma^0, \sigma^1, \sigma^2,\sigma^3\right\}, \quad 
\bar\sigma \equiv \left\{\sigma^0, -\sigma^1, -\sigma^2, -\sigma^3\right\}. \label{sigma}
\end{equation}
For both the Euclidean and Lorentzian cases, by definition the left and right chiral spinors $\psi_{\mathcal L}$ and $\psi_{\mathcal{R}}$
are defined to be eigenfunctions of the projections operators
\be
\psi_{\mathcal L} = \frac{1}{2}\left(1-\gamma^5\right)\psi_{\mathcal L}, 
\quad \psi_{\mathcal R} = \frac{1}{2}\left(1+\gamma^5\right)\psi_{\mathcal R}. \label{chirdef}
\ee
Since $\psi$ has four degrees of freedom, $\psi_{\mathcal L}$ and $\psi_{\mathcal R}$ have two degrees of freedom each (apart form color and flavor indices).

We are interested in the transformation properties of various spinor quadratic terms under Spin(4) and Spin(1,3)
transformations, accompanied by the corresponding SO(4) and SO(1,3) transformations
of vierbeins. In the Euclidean case we consider the following simultaneous pair of transformations:
\bea
\mbox{Euclidean:}
\quad \left\{\begin{array}{c}
\mathrm{SO(4)}:\quad
 e_{\mu}^F(x)\longrightarrow e'^{F}_{\mu}(x) = 
\left[\exp\left(-\frac{\ii}{2}\alpha_{AB}\Sigma^{AB}_{\mathrm E}\right)\right]_{~ \,G}^{F\,~}e^{G}_{\mu}(x),\\ \\
\mathrm{Spin(4)}:\quad \psi_{\alpha}(x)\longrightarrow\psi'_{\alpha}(x) = \left[\exp\left(-\frac{\ii}{2}\alpha_{AB}\sigma^{AB}_{\rm E}\right)\right]_{\alpha\,~}^{~\,\beta}\psi_{\beta}(x) 
\end{array} \right. \label{Etra}
\eea
where $\sigma^{AB}_{\rm E}$ stands for the generators of  the defining representation of Spin(4)
\begin{equation}
\sigma^{AB}_{\rm E} \equiv \frac{i\left[\gamma^A_{\rm E}, \gamma^B_{\rm E}\right]}{4},
\end{equation}
and $\Sigma_{AB}$ stand for the generators of the defining representation of SO(4). Correspondingly in the Lorentzian
case we are interested in the invariance under
\bea
\mbox{Lorentzian:}
\quad \left\{\begin{array}{c}
\mathrm{SO(1,3)}:\quad
 e_{\mu}^F(x)\longrightarrow e'^{F}_{\mu}(x) = 
\left[\exp\left(-\frac{\ii}{2}\alpha_{AB}\Sigma^{AB}_{\mathrm M}\right)\right]_{~ \,E}^{F\,~}e^{E}_{\mu}(x),\\ \\
\mathrm{Spin(1,3)}:\quad \psi_{\alpha}(x)\longrightarrow\psi'_{\alpha}(x) 
= \left[\exp\left(-\frac{\ii}{2}\alpha_{AB}\sigma^{AB}_{\rm M}\right)\right]_{\alpha\,~}^{~\,\beta}\psi_{\beta}(x) 
\end{array} \right. , \label{Ltra}
\eea
with  $\sigma^{AB}_{\rm M}$ generators of the defining representation of Spin(1,3)
\begin{equation}
\sigma_{AB}^{\rm M} \equiv \frac{i\left[\gamma_A^{\rm M}, \gamma_B^{\rm M}\right]}{4}, \label{Minkgen}
\end{equation}
and $\Sigma^{AB}_{\mathrm M}$ stand for generators of the defining representation of SO(1,3). In both formulas
\eqref{Etra} and \eqref{Ltra} we denote through $\alpha_{AB}$ six independent real parameters, and
$\alpha_{AB} = -\alpha_{BA}$, for $A,B=0,1,2,3$.

Apart from the original spinors, we also consider the charge conjugated spinors obtained by the action of
the charge conjugation operator $C$.  
 In particular in the Lorentzian case\footnote{We indicate complex conjugation by ${}^*$.} 
\be
C_{\mathrm M}\psi = -\ii\gamma^2_{\mathrm M}  \psi^*, \label{CM}
\ee
see for example~\cite[Eq.~3.145]{PeskinSchroeder},  
while in the Euclidean
\be
C_{\mathrm E}\psi =\ii\gamma^0_{\mathrm E}\gamma^2_{\mathrm E}  \psi^*. \label{CE}
\ee
In what follows it is convenient to use the following representation:
\be
C_{\mathrm E} = \hat{C}_{\rm E} \circ cc,
\ee
where $cc$ is complex conjugation and $\hat{C}_{\rm E} = \ii\gamma^0_{\mathrm E}\gamma^2_{\mathrm E}$ 
is a unitary matrix.  The spinors $C_{\mathrm E}\psi$ and $C_{\mathrm M}\psi$ transform as $\psi$
under Spin(4)  and Spin(1,3)  transformations respectively, but by the complex
conjugated representation under the action of the gauge group. Note, that the Lorentzian charge conjugation $C_{\rm M}$ changes chirality, i.e.\ it
maps left chiral spinor into right chiral spinor and vice versa, while the Euclidean charge conjugation $C_{\rm E}$ maps
left into left and right into right chiral spinors, i.e.\ it  preserves chirality.

In the Lorentzian case one introduces the kinetic and Dirac mass terms, which are invariant under \eqref{Ltra}:
\begin{equation}
\bar\psi\, \gamma^{A}_{\rm M}\, e^{\mu}_A \left(
\left[\nabla^{\rm{LC}}_{\mu}\right]^{\rm M} + \ii A_{\mu}\right)
\psi ,
\quad\bar\psi \psi,   \label{13inv}
\end{equation}
where $\bar\psi \equiv \psi^{\dagger} \gamma^0$ and $A_{\mu}$ is some vector field. Hereafter  $\nabla_{\mu}^{\rm{LC}}$ stands for the covariant derivative on the spinor bundle from the Levi-Civita spin-connection, which is different in the Euclidean and Lorentzian cases, see Appendix~\ref{appdetail}.    
The corresponding terms with the required Spin(4) invariance are:
\begin{equation}
\psi^{\dagger}\,\gamma^{A}_{\rm E}\, e^{\mu}_A \left(\left[\nabla^{\rm{LC}}_{\mu}\right]^{\rm E} + \ii A_{\mu}\right)
\psi,
\quad \psi^{\dagger} \psi . \label{4inv} 
\end{equation} 
Note that  the Majorana mass terms, built contracting  spinors with charge conjugated spinors, are invariant under both Spin(4) and Spin(1,3) actions, in particular:
\be
\underbrace{\left(C_{\mathrm E}\psi\right)^{\dagger}\psi}_{Spin(4)~\mathrm{inv}} 
= (-i\gamma^0_{\mathrm E}\gamma^2_{\mathrm E}\psi^*)^\dagger\psi
=\overline{(\gamma^2_{\mathrm M}\psi^*)}\psi
=-\underbrace{i\, \overline{\left(C_{\mathrm M}\psi\right)}\psi}_{Spin(1,3)~\mathrm{inv}} \label{doubleinv}
\ee

It is remarkable that, under the Wick rotation of the vierbeins $e_{\mu}^0 \rightarrow \ii e_{\mu}^0$, the ``rotationally"
invariant expression \eqref{4inv} \emph{does not} transform into Lorentz invariant structure \eqref{13inv}, unless
one inserts $\gamma^0$ by hand. We emphasize, that the Majorana mass terms \eqref{doubleinv} do not depend on vierbeins, and are both
Lorentz \eqref{13inv} and ``rotationally" \eqref{4inv} invariant \emph{without} any $\gamma^0$ insertion.

The Euclidean spectral action deals with the structures which are slightly different from the ones in~\eqref{4inv}.
Even after the removal of mirror fermions one has twice more independent spinors. In particular the kinetic and the Dirac mass terms are given by
\be
\left(C_{\rm E}\, \xi\right)^{\dagger}\gamma^{A}_{\rm E} \, e^{\mu}_A \left[\nabla^{\rm{LC}}_{\mu}\right]^{\rm E}
\psi , \quad
\left(C_{\rm E}\, \xi\right)^{\dagger} \psi,  \label{doubleEucl}
\ee
where $\xi$ and $\psi$ are independent spinors. These expressions are invariant under~\eqref{Etra}, and
transform under the Wick rotation of vierbeins $e_{\mu}^0\rightarrow \ii e^0_{\mu}$
into
\be
-\overline{\left(C_{\rm M}\, \xi\right)} \, \gamma^{A}_{\rm M} \, e^{\mu}_A \, \left[\nabla^{\rm{LC}}_{\mu}\right]^{\rm M}
\psi , \quad
\ii\overline{\left(C_{\rm M}\, \xi\right)} \psi,  \label{doubleEucl2}
\ee
which are invariant under \eqref{Ltra}. We emphasize, that the Spin(1,3) invariant expression \eqref{doubleEucl2} is obtained from the Spin(4) invariant one \eqref{doubleEucl} without any insertion of $\gamma^0$ by hand.
The extra spinorial degrees of freedom can be regarded as some sort of price to pay for such a simplification.

\subsection{Extra degrees of freedom. \label{extradegrees}}
In the algebraic approach to geometry, commutative and noncommutative manifolds are described  by real spectral triples, which
are defined by five entries $(\mathcal{A}, H, D, \gamma, J)$, where $\mathcal{A}$ is a (possibly noncommutative) algebra, represented on the Hilbert space $H$, and $D$ is an operator called a ``generalized Dirac operator" which acts on $H$, $\gamma$ and $J$ are operators called grading and real structure.
All five ingredients of the spectral triple must satisfy some relations known as ``axioms of noncommutative manifold", see~\cite{Connesreconstruction} 
for details. 
According to Connes reconstruction theorem
the ordinary commutative manifold $M$, with the spin structure, can be reconstructed from the infinite dimensional
``canonical" commutative spectral triple $(C_0^\infty(M), L^2(M,\mathcal{S}), \ii\slashed \nabla_{\rm E}^{\rm LC}, \gamma_5, C_{\rm E} )$, where $C_0^\infty(M)$ stands for the algebra of smooth  functions on $M$, with pointwise multiplication,  (vanishing at infinity in the non-compact case), $L^2(M,\mathcal{S})$ is the Hilbert space of square integrable Dirac (Spin(4)) spinors on $M$, the Dirac operator 
$\ii\slashed \nabla^{\rm LC}_{\rm E}\equiv \ii \gamma_{\rm E}^A \,e^{\mu}_A \,\left[\nabla_{\mu}^{\rm LC}\right]^{\rm E}$ is the usual one\footnote{This explains the terminology ``Dirac operator"
for arbitrary spectral triple.}
on a Riemannian spin manifold, the grading is given by chirality matrix $\gamma^5$, and the real structure is given by the Euclidean charge conjugation operator $C_{\rm E}$.

The spectral action approach to the Standard Model is based on seeing  it as an almost commutative geometry, which is defined by a product of an infinite dimensional  commutative ``canonical" spectral triple times a finite dimensional\footnote{{By finite a dimensional spectral triple we mean that the algebra and Hilbert spaces are finite dimensional vector spaces.} }noncommutative spectral triple $(A_F, H_F, D_F, \gamma_F, J_F)$ (for details see~\cite{Walterbook}).  To see the origin the fermionic quadrupling we focus our attention on the structure of $H$. 
The Hilbert space $H$ 
is given by the following tensor product 
\begin{equation}
H = L^2(M,\mathcal{S})\otimes H_F \label{Hprod}
\end{equation}
According to the construction the finite dimensional part $H_F$ is given by the direct sum
of left $H_L$, right $H_R$, anti-left $H_L^c$ and anti-right $H_R^c$ subspaces
\begin{equation}
H_F = H_L \oplus H_R \oplus H_L^c \oplus H_R^c, \label{HFsplit}
\end{equation}
Note the different notation $\mathcal L,R$ appearing in\eqref{chirdef} vs.\ $L,R$ in \eqref{HFsplit}, the former refers to a splitting in the Lorentzian indices, the latter in the gauge indices. In particular the subspaces $H_L$ and $H_R$ consist of the Dirac spinor multiplets which transform as left and right physical chiral multiplets under the action of the gauge group.
The corresponding dimensions $n = \mathrm{dim}(H_L) =  \mathrm{dim}(H_L^c)$, $m = \mathrm{dim}(H_R) =  \mathrm{dim}(H_R^c)$ equal to the number of left and right chiral fermions and take into account flavor and color indices in the physical model. 
These two numbers are not constrained and can be generally different.  For the standard model $n=m = 24$, (three colors of quarks   plus lepton, times two for ``up" and ``down" flavors, times three generations)  hence  ${\rm dim}\, H_F = 96$, while each spinor $\psi\in L^2(M,\mathcal{S})$ has four independent complex components. Therefore each element of $H$ is locally a vector valued functions with $4\cdot 96 = 384$ independent complex components. According to this construction, each chiral fermion of the SM and each chiral antifermion are present in the spectral action as independent Dirac spinors.
On the other side each physical chiral fermion (i.e.\ the field which appears in the Lagrangian) satisfies the relations~\eqref{chirdef}, i.e.\ is actually represented by a two component Weyl spinor
For example the subspace $H_L$ in \eqref{HFsplit},
which consists of spinors with (gauge) quantum numbers of left physical (Lorentzian) fermions, has both left $\left(H_L\right)_{\mathcal L}$ and right $\left(H_L\right)_{\mathcal R}$ chirality subspaces. This means that each left handed physical fermion  enters in $H_L$ together with its mirror partner, the spinor, which transforms under a gauge transformations as the original spinor, but has opposite chirality. In the following we will call this doubling of extra degrees of freedom ``mirror doubling". The other half of the quadrupling instead doubles the particle/antiparticle degree of freedom. We call this second doubling ``charge-conjugation doubling'', it will play a fundamental role in Sect.~\ref{se:genralpres}.  

\begin{remark}
Extra degrees of freedom also appear  in the Euclidean quantum field theory constructed by Osterwalder and Schrader~\cite{OsterwalderSchrader}. Their construction is rendered in an axiomatic manner directly introducing the Euclidean quantum Fock space\footnote{
Despite the mismatch of number of degrees of freedom per $\vec{k}$,
the Euclidean fermionic Fock space, introduced in \cite{OsterwalderSchrader}, does \emph{not} contain the Lorentzian physical Fock space as a subspace (in contrast to the bosonic construction). The only connection between Lorentzian
and Euclidean quantum field theories lies in the opportunity to obtain the Lorentzian Green's function via analytical
continuation of matrix elements.} and operators acting on it, while
Connes' spectral action approach deals with the Hilbert space of classical Euclidean fields. On the one hand, for each value $\vec{k}$ of the spatial momentum Lorentzian fermionic theory exhibits four
one-particle states (particle and antiparticle of two polarizations). On the other hand, in the Osterwalder-Schrader's construction there are infinitely many more states:
twice more polarizations, while each one particle state is also labeled by $k_0$, which varies continuously, so one deals with an ``infiniting" rather than with a doubling.
Despite some superficial similarities, the  extra degrees of freedom in the two approaches are  formally unrelated.  
\end{remark}

The mirror doubling problem was solved in~\cite{LMMS,AC2M2} with the introduction of the projected space $H_+$, defined as
\be
H_+=(H_L)_{\mathcal L}\oplus(H_R)_{\mathcal R}\oplus(H_L^c)_{\mathcal R}\oplus(H_R^c)_{\mathcal L}= P_+\,H,\quad P_+\equiv\frac{\mathbb I+\gamma_5 \otimes \gamma_F}2. \label{Hp}
\ee
where the grading $\gamma_F$ of the finite spectral triple is given by
\be
\gamma_F=\mathrm{diag}(\mathbb -1_{n},1_{m}, 1_{n},-1_{m}). \label{gF}
\ee
This projection satisfies the physical requirement that (Lorentzian) antiparticle have the opposite chirality than the corresponding particles. Alternative gradings are possible, see for example~\cite{DD14,DeepPurple}.

In~\cite{AC2M2} the following Euclidian action, \emph{free of mirror doubling}, was introduced:
\be
S_F= \frac{1}{2}\langle J\psi,D\psi\rangle \label{Jaction}.\ \ \psi\in H_+ 
\ee
where the real structure of the product spectral triple is given by
\be
J= C_{\mathrm E}\otimes J_F \label{TotalJ} \ee
with $C_{\mathrm E}$ introduced in \eqref{CE} and 

\begin{equation}
J_F = \left( \begin{array}{cccc}
0_{n\times n} & 0_{n\times m} & 1_{n\times n} & 0_{n\times m}\\
0_{m\times n} & 0_{m\times m} & 0_{m\times n} & 1_{m\times m} \\
1_{n\times n} & 0_{n\times m} & 0_{n\times n} & 0_{n\times m}\\
0_{m\times n} & 1_{m\times m} & 0_{m\times n} & 0_{m\times m} 
\end{array}
\right)\circ cc. \label{JFmatrix}
\end{equation} 
The standard model Lagrangian depends on $96$ complex functions, while corresponding expression~\eqref{Jaction} depends on $192$.  The action~\eqref{Jaction} reproduces correctly the Pfaffian, i.e.\ the functional integral over fermions, despite the fact that one still has twice the physical degrees of freedom.
In Sect.~\ref{se:rotatefermions}, we show how to perform the further reduction.

\begin{remark}Another useful fact was noted in~\cite{Barrett}. Starting with a fermionic action involving the whole  space $H$, written as usual \emph{with Lorentzian signature}, but imposing the 
following projections
\begin{equation}
J\psi^{\rm phys}  = \psi^{\rm phys}, \quad \gamma\psi^{\rm phys}  = \psi^{\rm phys}. \label{solbarr}
\end{equation}
These projections get rid of the unwanted states, but leave open the the definition of the Hilbert space, since the inner product is not positively defined and the bosonic spectral action cannot be defined in the same framework. In principle one may carry out Wick rotation
to Euclidean signature, and then compute the bosonic action, however this object would not represent the spectral triple
anymore, being not a pure ``bosonic spectral action". Such a projection would not be compatible with Euclidean signature.\end{remark}

\subsubsection*{Explicit form of the fermionic action. }
For further discussions we need a more detailed expression for the fermionic action \eqref{Jaction}.
The real structure $J$, given by \eqref{TotalJ}, acts on the subspace $H_+$ defined by \eqref{Hp} as:
\be
J H_+= C_{\mathrm E}(H_L^c)_{\mathcal R}\oplus 
             C_{\mathrm E}(H_R^c)_{\mathcal L}\oplus
             C_{\mathrm E}(H_L)_{\mathcal L}\oplus 
             C_{\mathrm E} (H_R)_{\mathcal R}
\ee
In this basis the Dirac operator is a four by four block matrix which looks like:
\be
D=\left[
\begin{array}{cccc}
\ii \slashed\nabla^{\rm E} & M_D & 0& 0\\
M_D^{\dagger} & \ii\slashed\nabla^{\rm E} & 0 & M_M^{\dagger}\\
0 & 0 &\ii \slashed\nabla^{\rm E} & M_D^*\\
0 & M_M & M_D^{\rm T} &\ii \slashed\nabla^{\rm E}
\end{array}
\right] \label{DConn}
\ee
where $M_D$ is a matrix containing the Dirac mass terms (Higgs fields, Yukawa couplings, etc.) and $M_M$ the one for Majorana mass terms\footnote{Although Majorana mass terms were originally
introduced in this context as constants \cite{AC2M2}, in later approaches they give rise to a scalar field 
\cite{Chamseddine:2012sw,Grand, Grand2, Grand3} which allows to match the experimentally observed Higgs mass with this formalism.} (which we consider only for right handed particles). Here and in the following we omit all internal indices for brevity.

Parametrizing a typical element $\psi \in H$ as 
\be
\psi = \left[
 \begin{array}{c}
\psi_L \\ \psi_R \\ \psi_L^c \\ \psi_R^c
\end{array}
\right],
\ee
where each entry is an independent Dirac spinor, we arrive at the following expression for the fermionic action
\bea
 S_F^{\rm E} &=& \frac{1}{2} \int d^4 x\sqrt{g^{\rm E}} \left[
 \begin{array}{c}
C_{\rm E} \left(\psi_L^c\right)_{\mathcal R} 
\\ C_{\rm E} \left(\psi_R^c\right)_{\mathcal L} 
\\ C_{\rm E} \left(\psi_L\right)_{\mathcal L} 
\\ C_{\rm E} \left(\psi_R\right)_{\mathcal R}
\end{array}
\right]^{\dagger}\left[
\begin{array}{cccc}
\ii \slashed\nabla^{\rm E} & M_D & 0& 0\\
M_D^{\dagger} & \ii\slashed\nabla^{\rm E} & 0 & M_M^{\dagger}\\
0 & 0 &\ii \slashed\nabla^{\rm E} & M_D^*\\
0 & M_M & M_D^{\rm T} &\ii \slashed\nabla^{\rm E}
\end{array}
\right] \left[
 \begin{array}{c}
\left(\psi_L\right)_{\mathcal L} \\ \left( \psi_R \right)_{\mathcal R} 
\\ \left(\psi_L^c\right)_{\mathcal R} \\ \left(\psi_R^c\right)_{\mathcal L}
\end{array}
\right] \nonumber\\
 &=& \int d^4 x\sqrt{g^{\rm E}} \left[
 \begin{array}{c}
C_{\rm E} \left(\psi_L^c\right)_{\mathcal R} 
\\ C_{\rm E} \left(\psi_R^c\right)_{\mathcal L} 
\end{array}
\right]^{\dagger}\left[
\begin{array}{cc}
\ii \slashed\nabla^{\rm E} & M_D \\
M_D^{\dagger} & \ii\slashed\nabla^{\rm E} 
\end{array}
\right] \left[
 \begin{array}{c}
\left(\psi_L\right)_{\mathcal L} \\ \left( \psi_R \right)_{\mathcal R} 
\end{array}
\right] \nonumber \\
&+& \frac{1}{2}\int d^4x \sqrt{g^{\rm E}}\left\{ \left[C_{\mathrm E}\left(\psi_R\right)_{\mathcal R}\right]^{\dagger}M_M \left(\psi_R\right)_{\mathcal R}
 + \left[C_{\mathrm E}\left(\psi_R^c\right)_{\mathcal L}\right]^{\dagger}M_M^{\dagger} \left(\psi_R^c\right)_{\mathcal L} \right\}, \label{STP}
\eea
where spinors with without ``c" are independent. 

\begin{remark}{The replacement of the complex conjugated spinor by the new variable (in fact the charge conjugation doubling) has also been introduced by van Nieuwenhuizen and Waldron~\cite{vanNieuwenhuizen:1996tv} independently of the spectral triple formalism. They  Wick rotated the Lorentzian quantum field theory to the Euclidean version in a way suitable for construction of the Euclidean supersymmetric theory. There are similarities,  in particular the Euclidean fermionic action of~\cite{vanNieuwenhuizen:1996tv} contains as many fermionic degrees of freedom as the NCG Euclidean fermionic action~\eqref{Jaction}. Nevertheless technically our approach and the one of ~\cite{vanNieuwenhuizen:1996tv} differ in the main aspects. While we Wick rotate just the vierbeins, van Nieuwenhuizen and Waldron transform the fields (fermionic and gauge).  In general the two approaches are different as well: NCG requires one more fermionic doubling i.e. the mirror doubling in order to construct the spectral triple and consequently to define the bosonic spectral action, while in the approach of  ~\cite{vanNieuwenhuizen:1996tv} there is no necessity to introduce mirror fermions.
 }
 \end{remark}


\section{Wick rotation for fermions \label{se:rotatefermions}}

In this section we present a general procedure to go from an Euclidean fermionic field theory to a Lorentzian one,
in a manner that is applicable to the formalism of noncommutative geometry.
Starting with the Euclidean fermionic action we will eventually arrive at a physical Lorentzian theory, free from doublings. To avoid cumbersome notations we will describe only the essentiality of the Lagrangian, leaving aside indices and irrelevant (in this context) features.

\subsection{General Prescription}
We will proceed in  two steps. 
The starting point is the  fermionic action \eqref{Jaction}, explicitly given by \eqref{STP}. This action is invariant under Spin(4) - SO(4)
transformations \eqref{Etra}.

\medskip

\noindent \textbf{Step 1.} \emph{Restoration of Lorentz invariance.} Perform the Wick rotation, given by \eqref{bosWickvier} i.e.\  we repeat the bosonic case:
\begin{equation}
\mbox{Wick rotation:} \quad - S_F^{\rm E} \left[\mbox{spinors}, e_{\mu}^A \right] 
\longrightarrow  \ii S_F^{\rm M\,doubled}\left[\mbox{spinors}, e_{\mu}^A \right]
\label{fWick}
\end{equation} 
After this step we will obtain the  fermionic action $S_F^{\rm M}$, invariant under Spin(1,3) - SO(1,3)
transformations \eqref{Ltra}  but still exhibiting the charge-conjugation doubling.
The spinors are still vectors in $H_+$, although there is no positive definite Spin(1,3)-invariant inner product on $H_+$ making it a Hilbert space).

\medskip

\noindent\textbf{Step 2.} \emph{Elimination of extra degrees of freedom.}
The charge-conjugation doubling consists in the presence in the fermionic Lagrangian (before and after step.\ 1) of spinors from  all four subspaces of $H_+$:
$\left(H_L^c\right)_{\mathcal R}$,  
$\left(H_R^{c}\right)_{\mathcal L}$, $\left(H_L\right)_{\mathcal L}$ and 
$\left(H_R\right)_{\mathcal R}$, while the physical Lagrangian depends on spinors just from the last two. 

We perform, after the Wick rotation \eqref{fWick}, the following identification
of the variables in the Lagrangian from subspaces $H_L^c$ and $H_R^{c}$ with the variables from $H_L$ and $H_R$:
\be
\begin{cases}
\;\left(\psi_L^{c}\right)_{\mathcal R} \in  \underbrace{ \left(H_L^c \right)_{\mathcal R}  }_{\subset H_+}
\quad\mbox{has to be identified with}\quad
 C_{\mathrm M}\left(\psi_L\right)_{\mathcal L} , \quad \left(\psi_L\right)_{\mathcal L}\in\underbrace{\left(H_L\right)_{\mathcal L}}_{\subset H_+} \\
\;\left(\psi_R^{c}\right)_{\mathcal L} \in  \smash{\underbrace{ \left(H_R^c \right)_{\mathcal L}  }_{\subset H_+}}
\quad\mbox{has to be identified with}\quad
 C_{\mathrm M}\left(\psi_R\right)_{\mathcal R} , \quad \left(\psi_R\right)_{\mathcal R}\in\smash{\underbrace{\left(H_R\right)_{\mathcal R}}_{\subset H_+}}
\rule{0pt}{16pt}
\end{cases} \label{fWick2}\vspace{14pt}
\ee
From a purely technical point of view, this step leads to the first formula of~\eqref{solbarr}, the same result of~\cite{Barrett}. Conceptually the difference is in the raison d'\^etre of this paper, namely our starting point is Euclidean.
As we show below, this recovers correct (real) Lorentzian Lagrangian.
Note, that only spinors, which belong to the subspaces
$\left(H_{L}\right)_{\mathcal L}$ and $\left(H_R\right)_{\mathcal R}$ appear in the final expression.

The procedure is self consistent, since under the Spin(1,3) and gauge transformation the quantities on the left and on the right side of the prescription~\eqref{fWick2} transform in the same way, and have the same
chirality.  
We stress that this procedure lies beyond the noncommutative geometry formalism. Lorentzian signature is in principle inconsistent with the formalism, and therefore Step~1 introduces new elements in the theory. Step~2, on the other side, is self consistent only if it is done \emph{after} Step~1. Indeed, under Spin(4) left and right
hand sides of~\eqref{fWick2} transform in different ways, therefore such an identification in all reference frames (invariant under~\eqref{Etra}) makes sense only if the spinors
of $H_+$ are Lorentzian.

\subsection{How the general prescription works. \label{se:genralpres}}
In this section, we show explicitly how the prescription \eqref{fWick} gives us a standard Lorentzian fermionic action, free of any doublings, starting from the Euclidean expression~\eqref{STP}. Since we will need the explicit dependence of the mass terms on spinor indices, we parametrize them as follows:
\be 
M_D = \gamma^5\otimes H,\quad M_M = \gamma^5\otimes \omega, \label{massstruct}
\ee
where the  matrix valued scalar fields $H$ and $\omega$ act on internal indices (gauge, flavor,  etc), not related with the spin structure. We omit all indices, apart from spinorial ones.

\subsubsection*{First step: restoration of Lorentzian signature}
The vierbeins $e_{\mu}^A$ enter in the fermionic action \eqref{STP}  via $\sqrt{g^{\rm E}}$ and $\slashed{\nabla}^{\rm E}$, which
is given by
\be
 \slashed{\nabla}^{\rm E} =  
 g^{\mu\nu}_{\rm E} e_{\mu}^A   \gamma_{A}^{\rm E} \nabla_{\nu}^{\rm E}. \label{QQ}
\ee
The covariant derivative in \eqref{QQ} has the following structure (we omit the unit matrix in flavor space for brevity):
\begin{eqnarray}
\nabla_{\nu}^{\rm E} &=& \left[\nabla_{\nu}^{ \rm LC}\right]^{\rm E} + i A_{\nu}, 
 \label{nablaEucl}
\end{eqnarray} 
where  $A_{\mu}$ is a  gauge connection. 
In Appendix~\ref{appdetail} we show the transformation:
\be
\sqrt{g^{\rm E}}\,\slashed{\nabla}^{\rm E}
\longrightarrow \sqrt{-g^{\rm M}}\,\slashed{\nabla}^{\rm M}, \label{qQq}
\ee 
where
\begin{equation}
\slashed{\nabla}^{\rm M} \equiv g^{\mu\nu}_{\rm M} e_{\mu}^A   \gamma_{A}^{\rm M} \nabla_{\nu}^{\rm M},
\end{equation}
and the Lorentzian covariant derivative is 
\begin{eqnarray}
\nabla_{\nu}^{\rm M} &=& \left[\nabla_{\nu}^{ \rm LC}\right]^{\rm M} + i A_{\nu},\label{nablaMink}
\end{eqnarray} 
and the gauge connection $A_{\mu}$ is the same as in the Euclidean case.

Substituting \eqref{QQQ} in \eqref{STP} we obtain:
\bea
 -S_{F}^{\rm E}&\rightarrow& -\int d^4 x\sqrt{-g^{\rm M}} \left[
 \begin{array}{c}
C_{\rm E} \left(\psi_L^c\right)_{\mathcal R} 
\\ C_{\rm E} \left(\psi_R^c\right)_{\mathcal L} 
\end{array}
\right]^{\dagger}\left[
\begin{array}{cc}
\ii \slashed\nabla^{\rm M} & \ii M_D \nonumber\\
\ii M_D^{\dagger} & \ii\slashed\nabla^{\rm M} 
\end{array}
\right] \left[
 \begin{array}{c}
\left(\psi_L\right)_{\mathcal L} \\ \left( \psi_R \right)_{\mathcal R} 
\end{array}
\right] \label{qq} \\
&&-\frac{\ii}{2}\int d^4x \sqrt{-g^{\rm M}}\left\{ \left[C_{\mathrm E}\left(\psi_R\right)_{\mathcal R}\right]^{\dagger}M_M \left(\psi_R\right)_{\mathcal R}
 + \left[C_{\mathrm E}\left(\psi_R^c\right)_{\mathcal L}\right]^{\dagger}M_M^{\dagger} \left(\psi_R^c\right)_{\mathcal L} \right\},
\eea

This action is invariant under the Lorentz transformation \eqref{Ltra}. In particular no 
modification of the inner product, like the insertion of $\gamma^0$, is needed. 
Using the identity $C_{\rm E}  = \ii \gamma^0 C_{\rm M}$
one can easily rewrite \eqref{qq} as:
\begin{align}
-S_{F}^{\rm E} \;\,\to\;\;\, & \ii\Bigg( \int d^4 x\sqrt{-g^{\rm M}} \overline{\left[
 \begin{array}{c}
C_{\rm M} \left(\psi_L^c\right)_{\mathcal R} 
\\ C_{\rm M} \left(\psi_R^c\right)_{\mathcal L} 
\end{array}
\right]}
\left[
\begin{array}{cc}
\ii \slashed\nabla^{\rm M} & \ii M_D \\
\ii M_D^{\dagger} & \ii\slashed\nabla^{\rm M} 
\end{array}
\right] \left[
 \begin{array}{c}
\left(\psi_L\right)_{\mathcal L} \\ \left( \psi_R \right)_{\mathcal R} 
\end{array}
\right]  \notag \\
& +\frac{1}{2} \int d^4x \sqrt{-g^{\rm M}}\left\{\ii \overline{\left[C_{\mathrm M}\left(\psi_R\right)_{\mathcal R}\right]}M_M \left(\psi_R\right)_{\mathcal R}
 + \ii \overline{\left[C_{\mathrm M}\left(\psi_R^c\right)_{\mathcal L}\right]} M_M^{\dagger} \left(\psi_R^c\right)_{\mathcal L} \right\} \Bigg), \label{qqq}
\end{align} 
which is manifestly Lorentz invariant.

\subsubsection*{Second step: elimination of extra degrees of freedom.}

The Lorentz invariant action coming from \eqref{qqq} contains extra degrees of freedom,
and is not acceptable as it is not real, since each quantity which carries the index ``c" is independent from the one which does not. Indeed the typical structure of the action for a single Dirac spinor
in flat space time reads (we do not write down mass terms for brevity.): 
\be
\int d^4 x \,\bar{\xi}\,\ii\slashed{\partial}^{\rm M}\psi \label{we_have}
\ee
(where $\xi$ and $\psi$ are independent), while the conventional one is given by:
\be
\int d^4 x\,\bar\psi\,\ii\slashed{\partial}^{\rm M}\psi. \label{must_be}
\ee 
Note that:
\begin{list}{$\bullet$}{\leftmargin=1.6em}
\item The classical system described by~\eqref{we_have} has a phase space twice bigger then needed for the description of Dirac fermions. At the classical level the number (per infinitesimal spatial volume) of physical degrees of freedom (particles and antiparticles) is half the dimensions of the phase
space after all the constraints\footnote{To discuss phase spaces one has to take into account the fact that both Lagrangians correspond to constrained Hamiltonian systems, and all conjugated momenta are not independent. See for example the discussion in~\cite{Das}.} are taken into account. The Dirac field describes four particles: two particles with different polarizations and the corresponding antiparticles, therefore the real dimension of the phase space per infinitesimal spatial volume must be 8, correctly reproduced by~\eqref{must_be}. On the other side, for (4.11) the dimension of the phase space per infinitesimal volume is equal to 16.

\item After canonical quantization of~\eqref{must_be} the operator
 $\hat\psi _{\alpha}^{\dagger}$  is not independent from $\hat\psi_{\alpha}$, but related via Hermitian conjugation with respect
 to the inner product in the Fock space.  Since there is no constraint $\psi = \xi$, direct application of the canonical quantization procedure to \eqref{we_have} must exhibit non coinciding operators $\hat{\psi}$ and $\hat{\xi}$ on the quantum space of states. This can cause pathologies e.g.\ non Hermitian Hamiltonian operator. 
Indeed, replacing the classical fields by operators in the classical Hamiltonian resulting from \eqref{we_have}, one would get the structure $-\ii\int d^3 x \, \hat{\xi}^{\dagger} \gamma^0_{\rm M}\left(\gamma^j_{\rm M} \partial_j\right) \hat{\psi}$, which is not formally selfadjoint. 

\item Our procedure for the elimination of the anti charge doubling
 is nothing but the imposition of this missing constraint on the classical fermionic phase space, thereby extracting its canonically quantizable part.
 
\item
However, the path integral is not sensitive to the charge-conjugation doubling, in particular
 the Pfaffian~\cite{AC2M2} is reproduced correctly (see~\eqref{grint}):
 \be
 \int[d \bar \psi][ d\psi] e^{\ii  \int d^4 x\, \bar\psi\,\ii\slashed{\partial}^{\rm M}\psi} = 
 \int[d \bar \xi ][ d\psi] e^{\ii  \int d^4 x\, \bar\xi\,\ii\slashed{\partial}^{\rm M}\psi} \label{dprop}.
 \ee
 
\item Although in the path integral approach the Green's functions which come from \eqref{we_have} are reproduced correctly, the correct identification of the Fock space is still necessary to understand the asymptotic states in scattering processes.
\end{list} 
The physical Lagrangian is given by~\eqref{must_be}. We eliminate the charge-conjugation doubling extra states with the prescription~\eqref{fWick2}.
Since
$
C_{\rm M}^2 = 1,
$
we obtain from~\eqref{qqq}:
\bea
S_{F}^{\rm M\, doubled}\longrightarrow  S_{F}^{\rm M} &=& \int d^4x \sqrt{-g^{\rm M}}\left\{ 
\overline{\left[
 \begin{array}{c}
\psi_{\mathcal L} \\  \psi_{\mathcal R} 
\end{array}
\right]}
\left[
\begin{array}{cc}
\ii \slashed\nabla^{\rm M} & \ii \gamma^5\otimes H \\
\ii \gamma^5\otimes H^{\dagger} & \ii\slashed\nabla^{\rm M} 
\end{array}
\right] \left[
 \begin{array}{c}
\psi_{\mathcal L} \\  \psi_{\mathcal R} 
\end{array}
\right] \right. \nonumber\\
 &&+ \left. \frac{1}{2} 
\left(\ii \overline{\left[C_{\mathrm M}\psi_{\mathcal R}\right]} \left(\gamma^5\otimes\omega\right) \,\psi_{\mathcal R}
 + \mbox{c.c.} \right)\right\}.  \label{qqqq}
\eea 
Because of the identification~\eqref{fWick2} the variables  $\left(\psi^c_R\right)_{\mathcal L}$ and $\left(\psi^c_L\right)_{\mathcal R}$ have disappeared from the action, and we are left with $\left(\psi_L\right)_{\mathcal L}$ and $\left(\psi_R\right)_{\mathcal R}$.
Since there is no risk of confusion anymore, hereafter we simplify the notations:
\be
\mbox{change of notations}:\quad 
\left(\psi_L\right)_{\mathcal L} \longrightarrow \psi_{\mathcal L}, \quad 
\left(\psi_R\right)_{\mathcal R} \longrightarrow \psi_{\mathcal R}, 
\ee
Following~\cite{Schucker}, we carry out a global axial transformation in order to recover the ``standard textbook"
form of the fermionic action. It is simple exercise using the (anti)commutation properties of the $\gamma$'s to show that, for an arbitrary $\alpha$, the kinetic term remains invariant under the following global axial
transformation
\begin{equation}
\psi_{\mathcal R,\mathcal L} \rightarrow e^{-i\alpha\gamma^5} \psi_{\mathcal R,\mathcal L}. \label{glax}
\end{equation}
Setting $\alpha = \pi/4$ one finds:
\begin{equation}
\psi_{\mathcal R,\mathcal L}\longrightarrow e^{-\frac{i\pi}{4}\gamma^5}\psi_{\mathcal R,\mathcal L}\
\Rightarrow\  \ii \, \overline{\psi_{\mathcal{L},\mathcal{R}}}
 \gamma^5  \left(\mbox{scalar}\right) \psi_{\mathcal R,\mathcal L} \longrightarrow - \overline{\psi_{\mathcal{L},\mathcal{R}}}
  \left(\mbox{scalar}\right) \psi_{\mathcal R,\mathcal L}. \label{glax2}
\end{equation}
It is easy to see that under the axial transformation~\eqref{glax} the conjugated 
spinors $C_{\rm M}\psi_{\mathcal{L}, \mathcal{R}}$
transform as the original ones $\psi_{\mathcal{L}, \mathcal{R}}$, therefore also for the Majorana mass terms we have
\begin{equation}
\ii \, \overline{\left[C_{\rm M}\left(\psi_{\mathcal{R}}\right)\right]}\, \omega\,
 \gamma^5 \, \psi_{\mathcal R} \longrightarrow - \, \overline{\left[C_{\rm M}\left(\psi_{\mathcal{R}}\right)\right]}\, \omega\,
  \, \psi_{\mathcal R} \label{glax3}
\end{equation}
Using \eqref{glax2} and \eqref{glax3} we can write down the fermionic action \eqref{qqqq}
with the new variables:
\bea
 S_{F}^{\rm M} &=& \int d^4 x\sqrt{-g^{\rm M}}
\left\{ 
\overline{\left(\psi_{\mathcal L}\right)}\, \ii \slashed\nabla^{\rm M}\psi_{\mathcal L}
+ \overline{\left(\psi_{\mathcal R}\right)}\, \ii \slashed\nabla^{\rm M}\psi_{\mathcal R}\right. \nonumber\\
&-&\left. \left[ \overline{\left(\psi_{\mathcal L}\right)}\,H\,\psi_{\mathcal R} 
+\frac{1}{2} \overline{\left[C_{\rm M}\left(\psi_{\mathcal R}\right)\right]}\,\omega\,\psi_{\mathcal R}
+ \mbox{c.c.}\right] 
\right\} \label{SFMfinal}
\eea
Care must be taken with global axial transformations when considering
path integrals. We show in  Appendix~\ref{apppathintegral} that the path integration must be performed \emph{after}
this axial transformation, in order to avoid gauge topological terms which come out from the axial anomaly and
modify the Green's functions. 

Let us rewrite the Lorentzian fermionic action in terms of two component Weyl spinors:
\be
\psi_{\mathcal L} = \left(\begin{array}{c}
\chi_L \\ 0
\end{array}\right),\quad \psi_{\mathcal R} = \left(\begin{array}{c}
0 \\ \chi_R
\end{array}\right), \label{chidef}
\ee
where $\chi_L$ and $\chi_R$ are two component Weyl spinors, which absorb all nonzero components
of $\psi_{\mathcal L}$ and $\psi_{\mathcal R}$ correspondingly.

Below we will use the following notations:  
\begin{equation}\begin{split}
\bar{\slashed{d}} \equiv g^{\mu\nu}_{\rm M} 
e_{\mu}^{A} \bar{\slashed{\sigma}}_{A}\nabla_{\nu}^{\rm Weyl \,L}, \\ 
{\slashed{d}} \equiv g^{\mu\nu}_{\rm M} 
e_{\mu}^{A} {\slashed{\sigma}}_{A}\nabla_{\nu}^{\rm Weyl \, R}.
\end{split}\end{equation}
where $\sigma$ and $\bar\sigma$ are defined in \eqref{sigma}, and
 $\nabla_{\nu}^{\rm Weyl \,L}$ and $\nabla_{\nu}^{\rm Weyl \,R}$ are  covariant derivativs on Weyl left
and right spinor bundles correspondingly:
\begin{equation}\begin{split}
\nabla_{\nu}^{\rm Weyl \,L} &= \left(\partial_{\mu} +i A_{\mu}\right)\otimes 1_2^{\rm W} - \frac{i}{2}
\left[\omega_{\nu}^{AB}\right]^{\rm M}\sigma^{\rm Weyl \,L}_{AB}  \\
\nabla_{\nu}^{\rm Weyl \,R} &= \left(\partial_{\mu} +i A_{\mu}\right)\otimes 1_2^{\rm W} - \frac{i}{2}
\left[\omega_{\nu}^{AB}\right]^{\rm M}\sigma^{\rm Weyl\, R}_{AB}
\end{split}\end{equation}
where $1_2^{\rm W}$ is a unity in Weyl spinor indexes, and $\sigma^{\rm Weyl \,L}_{AB}$ and $\sigma^{\rm Weyl \,R}_{AB}$ stand for the generators of left and right Weyl spinor representations of Spin(1,3), which are given by
\begin{equation}\begin{split}
\sigma_{jk}^{\rm Weyl\, L} &= \sigma_{jk}^{\rm Weyl\, R}  =  -\frac{i}{4}\left[\sigma_j, \sigma_k\right], \quad j,k =1,2,3,
 \\
\sigma_{j0}^{\rm Weyl\,L} &= - \sigma_{0j}^{\rm Weyl\,L}  =  \frac{i}{2}\sigma_j,\quad j=1,2,3, \\
\sigma_{j0}^{\rm Weyl\,R} &= - \sigma_{0j}^{\rm Weyl\,R}  =  -\frac{i}{2}\sigma_j,\quad j=1,2,3.
\end{split}\end{equation}

In terms of the two component spinors introduced by \eqref{chidef} the Lorentzian fermionic action 
reads
\begin{eqnarray}
S_F^{\rm M} & =&\int d^4 x \sqrt{-g^{\rm M}} \left\{
\underbrace{\chi_L^{\dagger}i 
\bar{\slashed{d}}
\chi_L + \chi_R^{\dagger}i 
\slashed{d}
\chi_R 
 }_{\mbox{ kinetic terms}}\,\,\,  - \underbrace{\left[ \chi_L^{\dagger}H\chi_R  + \chi_R^{\dagger}H^*\chi_L  \right]}_{\mbox{Dirac scalar-spinor couplings}}
 \right. \nonumber \\  &&\left.
  + \underbrace{\frac{1}{2} \left[
 i\chi_R^{\dagger}\sigma_2 \omega^* \chi_R^* - i\chi_R^T\sigma_2  \omega \chi_R 
   \right] 
 }_{\mbox{Majorana scalar-spinor couplings }}    \right\}.
\label{interm5}   
 \end{eqnarray}

\section{Conclusions}\label{sec:concl}
There are two themes which mingled in this paper: we discussed the Wick rotation of bosons and fermions from a Euclidean theory to a Lorentzian one, and the role of fermion doubling and its elimination. The most interesting result is the fact that these two issues are related, a relation that is probably even deeper than what presented here. 

First, the fermionic action~\eqref{Jaction} written with the real structure  $J$, which, as we explained, exhibits the charge-conjugation doubling, was introduced without any reference to Lorentz signature, and we have shown that the elegant vierbein Wick rotation procedure immediately recovers Lorentz invariance. In particular no  modification of the inner product ``by hand" is needed in this construction.
 This points to a role of the real structure $J$ also in this context.  
 
Second, we gave a prescription for the elimination  of  the remaining charge-conjugation doubling, thereby solving completely
the fermionic quadrupling problem. In particular, we have shown how one can arrive from the expression \eqref{STP} to the physically acceptable one \eqref{interm5} via the two steps prescription \eqref{fWick} and \eqref{fWick2}, where the former step is identical to the bosonic case, while the latter addresses peculiar features of fermionic theories.
Here we found another connection between extra degrees of freedom
and Lorentzian signature: we argued that the charge-conjugation doubling must be eliminated after the Wick rotation, i.e.~when the fermionic action is Spin(1,3) invariant.  An attempt to project out extra degrees of freedom in the Euclidean theory would immediately break the Spin(4) invariance.

The quadrupling of degrees of freedom is necessary to define the spectral action in its present formulation, which is Euclidean.  It does not correspond to physically observable\footnote{At least to low energy: in~\cite{Grand,Grand2,Grand3} there is a speculation of an higher energy ``pregeometric'' phase for which the quadrupling is necessary. From the results in this paper it follows that this hypothetical phase would also be Eucledian, along the lines of~\cite{HH}.} degrees of freedom. Half of the quadrupling is easily eliminated with a projection, while the charge-conjugation doubling, which cannot be projected out and creates troubles for the canonical quantization of a Lorentzian theory, allows for a simple Wick rotation.

While this paper solves the problem of the quadrupling, the solution, and its connection between Euclidean and Lorentzian theories may hint at more profound themes, yet to be discovered.

\appendix

\section{Remarks on the path integral \label{apppathintegral}}
Below we give a few comments relevant to path integrals. 
First we explain why the Pfaffian is not sensitive to the charge-conjugation doubling. Second we introduce
the correct measure in the Lorentzian path integral, remarking that one has to carry out the path integration \emph{after}
the field redefinition \eqref{glax2} to avoid the axial anomaly.

\subsubsection*{On the Grassmannian integration and the charge-conjugation doubling.}
It is interesting to explain how it happens that, although the fermionic action in~\cite{AC2M2} had extra degrees of freedom, the Pfaffian
was reproduced correctly. Technically the charge-conjugation doubling is a consequence of considering a spinor $\psi$ and it is
complex conjugated $\psi^*$ as independent variables $\psi$ and $\chi^*$ in the Lagrangians~\eqref{qqq} and~\eqref{qqqq}. 
An important algebraic fact  is the following.  No matter whether one integrates
over $\psi$ and $\psi^*$ or $\psi$ and $\chi^*$ (i.e.\ one considers twice more independent real variables), the resulting determinant is the same. This means that the anti-charge doubling has no effect on the Pfaffian (functional integral over fermions).
 In fact the following (somewhat counterintuitive) equality is valid:
\be
\int\prod_{n=1}^N [\dd\psi^*_n\dd\psi_n]\e^{\psi^*_jA_{jk}\psi_k} = 
\int\prod_{n=1}^N [\dd\chi^*_n\dd\psi_n]\e^{\chi^*_jA_{jk}\psi_k}=\det A \label{grint},
\ee
where $A$ is an arbitrary $N \times N$ matrix, and  $\psi_j$  and $\chi_j$ are truly independent \emph{complex}
Grassmanian variables.  Since this is important for our scopes, let us see it in detail.

The integration over a Grassmanian variable is equivalent to   taking  the derivative over it. In the complex case:
\be
\int \dd \psi_j 
= \vec{\partial}_{\psi_j}  \equiv \frac{1}{2}\vec{\partial}_{\xi_j} -\frac{\ii}{2} \vec{\partial}_{\eta_j}, \quad
\int \dd \psi^*_j = \vec{\partial}_{\psi^*_j}  \equiv \frac{1}{2}\vec{\partial}_{\xi_j} +\frac{\ii}{2} \vec{\partial}_{\eta_j},  \quad
\int \dd \chi^*_j 
= \vec{\partial}_{\eta_j}  \equiv \frac{1}{2}\vec{\partial}_{\theta_j} +\frac{\ii}{2} \vec{\partial}_{\lambda_j},
\ee
where $\psi_j = \xi_j + \ii \eta_j$, $\chi_j = \theta_j + \ii \lambda_j$,  and $ \xi_j$, $\eta_j$, $\theta_j$ and $\lambda_j$
are real fields.
The anticommutator of any pair of variables vanishes.

We emphasize, that the former integrand in \eqref{grint} depends on $2N$ real Grassmanian variables
while the latter integrand on $4N$ independent real Grassmanian variables. The integration rule, however,
leads to the same answer.
Indeed, although $\psi_j$ and $\psi_j^*$, being mutually complex conjugated are \emph{not} independent,
when one carries out the integration (= takes derivative) over them, they can be considered as independent
variables, since
\be
\partial_{\psi_j} \psi^* = \left(\frac{1}{2}\vec{\partial}_{\xi_j} -\frac{\ii}{2} \vec{\partial}_{\eta_j}\right)
\left(\xi_j - \ii\eta_j\right) = 0,
\ee
and
\be
\partial_{\psi_j^*} \psi = \left(\frac{1}{2}\vec{\partial}_{\xi_j} +\frac{\ii}{2} \vec{\partial}_{\eta_j}\right)
\left(\xi_j + \ii\eta_j\right) = 0.
\ee

\subsubsection*{On the correct measure in the path integral}
Below we explain, that the path integral over fermions has to be taken \emph{after} the global axial transformation
\eqref{glax2}, or more precisely the variables $\psi^{{\rm old}}$, which enter in the ``almost final" Minkowskian fermionic 
action $S_{F}^{\rm old}$ (given by \eqref{qqqq}) and
the variables $\psi^{{\rm new}} = e^{+\frac{i\pi}{4}\gamma^5}\psi^{\rm old}$, which enter  in the ``final" fermionic actions $S_{F}^{\rm new}$ (given \eqref{SFMfinal})  are not equivalent, since they lead to different Green's functions. For arbitrary
composite operator $\mathcal{O}$  which involves fields, coupled to fermions (directly or via quantum corrections) one obtains: 
\begin{align}
 \langle T \mathcal{O}\rangle_{\rm old\, fields} &\equiv \frac{
\int[d\mathcal B] [d\bar\psi^{\rm old}][d\psi^{\rm old}] \mathcal{O}
e^{\ii S_F^{\rm M\, old} +\ii S_{\rm bos} }}
{\int[d\mathcal B] [d\bar\psi^{\rm old}][d\psi^{\rm old}] 
e^{\ii S_F^{\rm M\, old} +\ii S_{\rm bos} }} = \frac{
\int[d\mathcal B] [d\bar\psi^{\rm new}][d\psi^{\rm new}] \mathcal{O}
e^{\ii S_F^{\rm M\, new} +\ii \tilde{S}_{\rm bos} }}
{\int[d\mathcal B] [d\bar\psi^{\rm new}][d\psi^{\rm new}] 
e^{\ii S_F^{\rm M\, old} +\ii \tilde{S}_{\rm bos} }} \notag \\
 & \neq \frac{
\int[d\mathcal B] [d\bar\psi^{\rm new}][d\psi^{\rm new}] \mathcal{O}
e^{\ii S_F^{\rm M\, new} +\ii S_{\rm bos} }}
{\int[d\mathcal B] [d\bar\psi^{\rm old}][d\psi^{\rm new}] 
e^{\ii S_F^{\rm M\, new} +\ii S_{\rm bos} }} \equiv \langle T \mathcal{O} \rangle_{\rm new\, fields}. \label{nonequiv}
\end{align}
where $T$ stands for time ordering, and $\mathcal{B}$ for bosonic measure. The change of the bosonic action 
\be
S_{\rm gauge}\rightarrow \tilde{S}_{\rm gauge} \equiv S_{\rm gauge} + (\rm{const})\, \epsilon^{\mu\nu\alpha\beta} \mathcal{F}_{\mu\nu} \mathcal{F}_{\alpha\beta},
\ee
where the tensor $ \mathcal{F}_{\mu\nu}$  corresponds to nonabelian gauge connection $\mathcal{A_{\mu}}$, 
came out from nontrivial Jacobian of the global axial transformation $\psi^{\rm old} \longrightarrow \psi^{\rm new}$. 
Indeed, although the ``old" action \eqref{qqqq} transforms into the ``new" one \eqref{SFMfinal} under the transformation 
$\psi^{\rm old}\longrightarrow\psi^{\rm new}$,  the fermionic measure $[d\bar\psi][d\psi]$ does not. This phenomenon is the so-called axial anomaly, see~\cite{Fujikawabook}: a gauge invariant regularizations of the functional integral over fermions introduces a dependence of the regularized measure on the gauge fields.
The Jacobian in flat space-time reads:
\be
\exp{\left(\ii(\rm{const})\int d^4 x \, \epsilon^{\mu\nu\alpha\beta} \mathcal{F}_{\mu\nu} \mathcal{F}_{\alpha\beta} \right) }, \label{tf}
\ee
When the nonabelian gauge field $\mathcal{A}_{\mu}$ has nontrivial Pontryagin number, the Jacobian is different from one.  
Taking the functional integral over the gauge field $\mathcal{A}_{\mu}$, various configurations
with nontrivial Pontryagin index give different contributions to the path integral, hence the inequality in~\eqref{nonequiv}. 
For example, setting $\mathcal{O} =\mathcal{A}_{\mu}(x) \mathcal{A}_{\nu}(y)$, we obtain different
full propagators for the gauge field in the ``new'' or ``old'' variables.
In order to work with the standard fermionic action \eqref{SFMfinal} and with the standard bosonic spectral
action $S_{\rm bos}$ without the topological term~\eqref{tf}, 
one has to postulate that the functional integration is done \emph{after} the global axial transformation \eqref{glax2} i.e.\
over the new variables $\psi^{\rm new}$.

\section{\hspace*{-8pt}\mbox{Notations and conventions for the gravitational sector}\label{appnotation}}

Throughout this paper we use the following notations:\\
Riemann tensor 
\begin{equation}
R^{\mu}_{~\nu\rho\sigma}\left[g_{\mu\nu}\right] = \partial_{\sigma}\Gamma^{\mu}_{\nu\rho} - \partial_{\rho}\Gamma^{\mu}_{\nu\sigma}
+ \Gamma^{\lambda}_{\nu\rho}\Gamma^{\mu}_{\lambda\sigma} - \Gamma^{\lambda}_{\nu\sigma}\Gamma^{\mu}_{\lambda\rho} \label{RMN}
\end{equation}
Ricci tensor 

\begin{equation}
R_{\mu\nu}\left[g_{\mu\nu}\right] = R^{\sigma}_{~\mu\sigma\nu} =   \partial_{\nu}\Gamma^{\sigma}_{\mu\sigma} -\partial_{\sigma}\Gamma^{\sigma}_{\mu\nu}+
\Gamma^{\lambda}_{\mu\sigma}\Gamma^{\sigma}_{\lambda\nu}
- \Gamma^{\lambda}_{\mu\nu}\Gamma^{\sigma}_{\lambda\sigma}  \label{RICCI}
\end{equation}
Scalar curvature 
\begin{equation}
R\left[g_{\mu\nu}\right] = g^{\mu\nu}\left\{  \partial_{\nu}\Gamma^{\sigma}_{\mu\sigma} - \partial_{\sigma}\Gamma^{\sigma}_{\mu\nu}
 + \Gamma^{\lambda}_{\mu\sigma}\Gamma^{\sigma}_{\lambda\nu}- \Gamma^{\lambda}_{\mu\nu}\Gamma^{\sigma}_{\lambda\sigma} \right\} \label{RS}
\end{equation}
with the Christoffel symbols of the second kind 
\begin{equation}
\Gamma_{\nu\rho}^{\mu}\left[g_{\mu\nu}\right]\equiv\frac{1}{2}\,g^{\mu\lambda}\left(\partial_{\rho}g_{\lambda\nu} + \partial_{\nu}g_{\lambda\rho} - \partial_{\lambda}g_{\nu\rho}\right) \label{Christ}
\end{equation}
Note also the identity
\be
C_{\mu\nu\alpha\beta}\left[g_{\mu\nu}\right] C^{\mu\nu\alpha\beta}\left[g_{\mu\nu}\right]
\equiv R_{\mu\nu\alpha\beta}\left[g_{\mu\nu}\right] R^{\mu\nu\alpha\beta}\left[g_{\mu\nu}\right]
- 2 R_{\mu\nu}\left[g_{\mu\nu}\right] R^{\mu\nu}\left[g_{\mu\nu}\right]
+\frac{1}{3}R^2\left[g_{\mu\nu}\right]  \label{W2}
\ee

\section{Derivation of   \eqref{qQq}\label{appdetail} }
In this appendix we derive the formula \eqref{qQq}.
The vierbeins enter in $\nabla_{\mu}^{\rm E}$ via $\left[\nabla_{\mu}^{\rm LC }\right]^{\rm E}$, 
therefore one has to show that, under the rotation~\eqref{fWick}, 
the covariant derivative $\left[\nabla_{\mu}^{\rm LC }\right]^{\rm E}$,
considered as a function of the vierbeins, will transform into $\left[\nabla_{\mu}^{\rm LC }\right]^{\rm M}$.
We need the explicit expression for the covariant derivative  
$\left[\nabla_{\mu}^{\rm LC }\right]^{\rm E}$:
\begin{equation}
 \left[\nabla_{\mu}^{\rm LC }\right]^{\rm E} = \partial_{\mu}\otimes 1_4^{\rm s}  - \frac{i}{2}\omega_{\mu}^{\rm E}, 
\end{equation}
where $1_4^{\rm s}$ is a unity in spinor indexes, and 
the Euclidean spin connection is given by
\begin{equation}
\omega_{\mu}^{\rm E}\equiv \left[\omega_{\mu}^{AB}\right]^{\rm E}  \sigma_{AB}^{\rm E}, \label{oeucl1}
\end{equation}
with
\begin{equation}
\left[\omega_{\mu}^{AB}\right]^{\rm E} \equiv e^{A}_{\nu} g^{\nu\alpha}_{\rm E}   \partial_{\mu} e^{B}_{\alpha} 
+ e^{A}_{\nu} \left[\Gamma^{\nu}_{\mu\sigma}\right]^{\rm E} e^{B}_{\beta} g^{\beta\sigma}_{\rm E}, \label{oeucl}
\end{equation}
where $\left[\Gamma^{\nu}_{\mu\sigma}\right]^{\rm E}$ is expressed via $g_{\mu\nu}^{\rm E}$ according to
\eqref{Christ} and $g_{\mu\nu}^{\rm E}$ depends on vierbeins via \eqref{geucl}, i.e.\ \eqref{oeucl} is just a function
of the vierbeins. In order to prove, that for \eqref{fWick}
\begin{equation}
{\rm Wick:} \quad \left[\nabla_{\mu}^{\rm LC }\right]^{\rm E} \longrightarrow \quad \left[\nabla_{\mu}^{\rm LC }\right]^{\rm M} \equiv
\partial_{\mu}\otimes 1_4^{\rm s}  - \frac{i}{2}\omega_{\mu}^{\rm M}, \label{nablaspin}
\end{equation}
one has to show that
\begin{equation}
{\rm Wick:} \quad  \omega_{\mu}^{\rm E} \longrightarrow \omega_{\mu}^{\rm M},
\end{equation}
where the latter is given by
\begin{equation}
\omega_{\mu}^{\rm M}\equiv \left[\omega_{\mu}^{AB}\right]^{\rm M}  \sigma_{AB}^{\rm M}. \label{omink1}
\end{equation}

The spin connection coefficients in the Lorentzian case are 
\begin{equation}
\left[\omega_{\mu}^{AB}\right]^{\rm M} \equiv e^{A}_{\nu} g^{\nu\alpha}_{\rm M}   \partial_{\mu} e^{B}_{\alpha} 
+ e^{A}_{\nu} \left[\Gamma^{\nu}_{\mu\sigma}\right]^{\rm M} e^{B}_{\beta} g^{\beta\sigma}_{\rm M}, \label{omink}
\end{equation}
where again $\left[\Gamma^{\nu}_{\mu\sigma}\right]^{\rm M}$ is expressed via $g_{\mu\nu}^{\rm M}$ according to
\eqref{Christ} and $g_{\mu\nu}^{\rm M}$ depends on vierbeins via \eqref{gmink}, i.e.\ \eqref{omink} is again just a function
of the vierbeins, different from \eqref{oeucl}.
After we introduced all notation, one can rewrite \eqref{oeucl1}:
\begin{eqnarray}
\omega_{\mu}^{\rm E} &=& \sum_{k,j=1}^3 \left[\omega_{\mu}^{kj}\right]^{\rm E}  \sigma_{kj}^{\rm E}
+2 \sum_{j=1}^3 \left[\omega_{\mu}^{0j}\right]^{\rm E}  \sigma_{0j}^{\rm E}   \nonumber \\
&=& \sum_{k,j=1}^3 \left[\omega_{\mu}^{kj}\right]^{\rm E}  \left(-\sigma_{kj}^{\rm M}\right)
+2 \sum_{j=1}^3 \left[\omega_{\mu}^{0j}\right]^{\rm E}  \left(i\sigma_{0j}^{\rm M}\right) \label{interm}
\end{eqnarray}

Now we are prepared for the final stroke: the Wick rotation~\eqref{fWick}.  Since both indices $A$ and $B$ in \eqref{oeucl} are carried by  vierbeins, and since under~\eqref{fWick}
the metric tensor $g_{\mu\nu}^{\rm E} \longrightarrow -g_{\mu\nu}^{\rm M} $ and 
$\left[\Gamma_{\mu\nu}^{\lambda}\right]^{\rm E} \longrightarrow  \left[\Gamma_{\mu\nu}^{\lambda}\right]^{\rm M} $,
we immediately obtain:
\begin{equation}
\mbox{Wick:}\quad\quad\left\{ \begin{array}{c} 
\left[\omega_{\mu}^{0j}\right]^{\rm E}  \longrightarrow -i\left[\omega_{\mu}^{0j}\right]^{\rm M}, \quad j = 1,2,3 
 \\  \left[\omega_{\mu}^{kj}\right]^{\rm E}  \longrightarrow -\left[\omega_{\mu}^{kj}\right]^{\rm M}, \quad k,j = 1,2,3 
 \end{array}
\right. \label{interm2}
\end{equation}
Substituting \eqref{interm2} into \eqref{interm} we see that
the spin connection $\omega_{\mu}$ transforms in the proper way:
\begin{equation}
\mbox{Wick}\quad \omega_{\mu}^{\rm E} \longrightarrow 
\sum_{k,j=1}^3 \left[\omega_{\mu}^{kj}\right]^{\rm M}  \sigma_{kj}^{\rm M}
+2 \sum_{j=1}^3 \left[\omega_{\mu}^{0j}\right]^{\rm M}  \sigma_{0j}^{\rm M} \equiv \omega_{\mu}^{\rm M},
\end{equation}
therefore the equality \eqref{nablaspin} is proven.

Expressing $\gamma_A^{\rm E}$ via $\gamma_A^{\rm M}$ according to \eqref{gammaEM} and using:
\begin{equation}
ie_{\mu}^A   \gamma_{A}^{\rm E} = i e_{\mu}^0   \gamma_{0}^{\rm E} + i e_{\mu}^j   \gamma_{j}^{\rm E}
 = i e_{\mu}^0   \gamma_{0}^{\rm M} -  e_{\mu}^j   \gamma_{j}^{\rm M}
	\xrightarrow[\mathrm{Wick}]{}
 - e_{\mu}^A   \gamma_{A}^{\rm M},
\end{equation}
finally we arrive at the following law of transformation of $\slashed{\nabla}$:
\be
\mbox{Wick:}\quad \ii\slashed{\nabla}^{\rm E} \longrightarrow \slashed\nabla^{\rm M}
\quad\mbox{or}\quad\sqrt{g^{\rm E}}\,\,\ii\slashed{\nabla}^{\rm E} \rightarrow \quad\sqrt{- g^{\rm M}}\,\,\ii\slashed{\nabla}^{\rm M}.\label{QQQ}
\ee

\section*{Acknowledgments}
We would like of thank D.~Vassilievich for fruitful discussions. This article is based upon work from COST Action MP1405 QSPACE, supported by COST (European Cooperation in Science and Technology). FL and FDA are partially supported by INFN Iniziativa Specifica GeoSymQFT.  FL is partially supported by CUR Generalitat de Catalunya under projects FPA2013-46570 and 2014~SGR~104, MDM-2014-0369 of ICCUB (Unidad de Excelencia `Maria de Maeztu'). MAK is supported by the FAPESP process 2015/05120-0.

\end{document}